\documentstyle[12pt]{article}

\def\Diff{\hbox{\rm Diff}}
\def\S1{\hbox{\rm S$^1$}}
\def\Vect{\hbox{\rm Vect}}

\def\pds#1,#2{\langle #1\mid #2\rangle} 
\def\f#1,#2,#3{#1\colon#2\to#3} 

\def\hfl#1{{\buildrel{#1}\over{\hbox to
12mm{\rightarrowfill}}}}

\begin{document}

\title{Extensions of Virasoro group and Virasoro algebra by
modules of tensor densities on $S^1$}

\author{Valentin OVSIENKO
\thanks{C.N.R.S., Centre de Physique Th\'eorique, Luminy-Case 907, F-13288
Marseille Cedex 9, France}, Claude ROGER\thanks{URA 746 /Institut de
Math\'ematiques et d'Informatiques/, Universit\'e Claude Bernard - Lyon I, 43
bd. du 11 Novembre 1918, 69622 Villeurbanne Cedex, France}}

\maketitle

\abstract{We classify non-trivial (non-central) extensions of the group
$\Diff ^+(S^1)$ of all diffeomorphisms of the circle preserving its
orientation and of the Lie algebra $\Vect (S^1)$ of vector
fields on $S^1$, by the modules of tensor densities on $S^1$. The result is:
4 non-trivial extensions of $\Diff ^+(S^1)$ and 7 non-trivial extensions of
$\Vect (S^1)$. Analogous results hold for the Virasoro group and the Virasoro
algebra. We also classify central extensions of constructed Lie algebras}

\vfill\eject

\section{introduction}

The Lie group $\Diff ^+(S^1)$ of all orientation preserving
diffeomorphisms of the circle, has a unique (up
to isomorphism) non-trivial central extension, so-called
{\it Bott-Virasoro group}. It is defined by the Thurston-Bott cocycle [B]:

$$B(\Phi ,\Psi )=\int_{S^1}log((\Phi \circ \Psi )')dlog(\Psi ')$$
where $\Phi ,\Psi \in \Diff ^+(S^1)$, the function $\Phi '=\frac{d\Phi
(x)}{dx}$ being well defined on $S^1$.

The corresponding Lie algebra is called the {\it Virasoro algebra}. It is
given by the unique (up to isomorphism) non-trivial central extension of
the Lie algebra $\Vect (S^1)$ of all vector fields on the circle. This central
extension is defined by the Gelfand-Fuchs cocycle [GF 1]:

$$\omega (f,g)=\int_{S^1}\left| \begin{array}{cc}
f^{\prime}&g^{\prime} \\
f^{\prime\prime}&g^{\prime\prime}\\
\end{array}\right|dx$$

In this paper we study (non-central) extensions of the group $\Diff ^+(S^1)$
and the Lie algebra $\Vect (S^1)$ by the modules of tensor
densities on $S^1$.

\vskip 0,3cm

Let ${\cal F}_\lambda $ be the space of all tensor densities on $S^1$ of
degree $\lambda $:

$$a=a(x)(dx)^{\lambda }$$
This space has natural structures of $\Diff (S^1)$ and $\Vect (S^1)$-module.
The $\Diff (S^1)$-action on ${\cal F}_\lambda $ is given by
$$\Phi ^*a=a(\Phi )(\Phi ')^{\lambda }$$
The Lie algebra $\Vect (S^1)$ acts on this space by the Lie derivative: let
$f=f(x)d/dx$ be a vector field, then
 $$L_fa=(fa'+\lambda f'a)(dx)^{\lambda }$$

We consider the problem of classification of
all non-trivial extensions

\begin{eqnarray}
0\rightarrow {\cal F}_\lambda \rightarrow G_\lambda \rightarrow
\Diff ^+(S^1)\rightarrow 0
\end{eqnarray}
of the group $\Diff ^+(S^1)$ by
$\Diff ^+(S^1)$- modules ${\cal F}_\lambda $.

In other words, we are looking for group structures on
$\Diff ^+(S^1)\times {\cal F}_\lambda $ given by
associative product of the following form:

$$(\Phi ,a)(\Psi ,b)=(\Phi \circ \Psi ,\; b+\Psi ^*a+B_{\lambda }(\Phi ,\Psi
))$$
The expression $B_{\lambda }(\Phi ,\Psi )\in {\cal F}_\lambda $ satisfyies
the  condition:
$$B_{\lambda }(\Phi ,\Psi \circ \Xi )+ B_{\lambda }(\Psi ,\Xi )=
B_{\lambda }(\Phi \circ \Psi ,\Xi )+\Xi ^*B_{\lambda }(\Phi ,\Psi )$$
which means that $B_{\lambda }(\Phi ,\Psi )$ is a 2-cocycle on $\Diff S^1$
with values in ${\cal F}_\lambda $. If $B_{\lambda }=0$ then the group
$G_{\lambda }$ is
called the  semi-direct product: $G_{\lambda }=\Diff ^+(S^1)\triangleright
{\cal F}_\lambda $.

 The extension (1) is {\it
non-trivial} if the Lie group $G_{\lambda }$ is not isomorphic to
$\Diff S^1\triangleright {\cal
F}_\lambda $. The cocycle
$B_{\lambda }$ in this case, represents a non-trivial cohomology class of the
groupe $H^2(\Diff ^+(S^1);{\cal F}_\lambda )$. The classification
problem for the extensions (1) is equivalent to the problem
of computing this cohomology group.

We calculate the group $H^2_c(\Diff ^+(S^1);{\cal F}_\lambda )$ of
differentiable cohomology in Van-Est's sense. It means, that we classify all
the extensions given by differentiable cocycles. We find four non-isomorphic
infinite-dimensional Lie groups. We give explicit formulae for non-trivial
cocycles on $\Diff ^+(S^1)$.

We also consider non-trivial extensions of the Lie algebra $\Vect (S^1)$:

\begin{eqnarray}
0\rightarrow {\cal F}_\lambda \rightarrow g_\lambda
\rightarrow \Vect (S^1)\rightarrow 0
\end{eqnarray}
One obtains the classification of these extensions as a corollary of
some general theorems in the cohomology theory of infinite-dimensional Lie
algebras.

 On classifying the extensions (2), one finds a series of seven Lie algebras.
They are defined on the space $\Vect (S^1)\oplus {\cal F}_\lambda $. The
commutator is given by:

$$
[(f,a)(g,b)]=([f,g],L_fb-L_ga+c(f,g))
$$
where $c$ is a 2-cocycle on $\Vect (S^1)$ with values in ${\cal F}_\lambda $.

 We classify non-trivial central extensions of all the Lie algebras given
by the extensions (2). Some of these Lie algebras have already been
considered in the mathematical literature, some of them are probably new.

All
the Lie groups and the Lie algebras defined by the extensions (1) and
(2) seem to be interesting generalisations of the Virasoro group and algebra.
Their representations, coadjoint orbits etc. appear be interesting
subjects, and deserve further study.

\section{Main theorems}
Let us formulate main results of this paper
\subsection{Extensions of the group $\Diff ^+(S^1)$}

The following theorem gives a classification of non-trivial extensions (1)
up to an isomorphism.

\proclaim Theorem 1. One has for cohomology of the group $\Diff ^+(S^1)$ :
$$
H^2_c(\Diff ^+(S^1);{\cal F}_\lambda )=\left\{\matrix{
{\bf R},\; \lambda =0,1,2,5,7  \cr
0,\; \lambda \not = 0,1,2,5,7  \cr
}\right.
$$\par

In other words, there exists a unique (modulo isomorphism)
non-trivial extension of $\Diff ^+(S^1)$ by the module ${\cal F}_\lambda $ for
each value: $\lambda =0,1,2,5,7$. If $\lambda \not =0,1,2,5,7$, there is no
non-trivial extensions.
\vskip 0,3cm

Let us describe here the 2-cocycles

$$B_\lambda :\Diff ^+(S^1)\times \Diff ^+(S^1)\rightarrow {\cal F}_\lambda$$
which generate all possible non-trivial cohomology classes.

First of all, recall that the following mappings:

\begin{eqnarray}
l:\Phi  &\mapsto &log(\Phi^{\prime}(x))   \cr
dl:\Phi &\mapsto &d\;log(\Phi^{\prime}(x))=\frac{\Phi^{\prime \prime}}{\Phi
^{\prime}}dx  \cr S:\Phi  &\mapsto &\left[\frac{\Phi
^{\prime\prime\prime}}{\Phi
^{\prime}}-\frac{3}{2}\left(\frac{\Phi
^{\prime\prime}}{\Phi^{\prime}}\right)^2\right](dx)^2
\end{eqnarray}
define 1-cocycles on $\Diff ^+(S^1)$ with values in
${\cal F}_0,{\cal F}_1,{\cal F}_2$, correspondingly. They represent unique
(modulo coboundaries) non-trivial classes of the cohomology groups:
$H^1(\Diff ^+(S^1);{\cal F}_\lambda ), \lambda =0,1,2$. The cocycle $S$ is
called the {\it Schwarzian derivative} , $dl$ is called the logarithmic
derivative.
\vskip 0,3cm

To construct non-trivial cocycles on $\Diff ^+(S^1)$, we shall use the
following version of the general {\it cap-product} (see e.i. [Br]) on group
coccycles. Consider a Lie group $G$. Let: $U,\;V,\;W$ be $G$-modules;
$u:G\rightarrow U,\;v:G\rightarrow V$ be 1-cocycles on $G$; $\langle
.,.\rangle :U\otimes V\rightarrow W$ be G-invariant bilinear mapping. Then
$$u\cap v(g,h)=\langle h(u(g)),\;v(h)\rangle $$
 is a 2-cocycle on $G$
with values in $W$.
\vskip 0,3cm

We are going to use the following invariant  bilinear mapping on the spaces
${\cal F}_\lambda$.

a) The product of tensor densities : ${\cal F}_\lambda
\otimes {\cal F}_\mu \rightarrow {\cal F}_{\lambda +\mu }$
$$a(x)(dx)^{\lambda }\otimes b(x)(dx)^{\mu }\mapsto a(x)b(x)(dx)^{\lambda
+\mu }.$$

b) The Poisson
bracket : ${\cal F}_\lambda
\otimes {\cal F}_\mu \rightarrow {\cal F}_{\lambda +\mu +1}$  $$\{
a(x)(dx)^{\lambda },b(x)(dx)^{\mu }\}=(\lambda a(x)b'(x)-\mu
a'(x)b(x))(dx)^{\lambda +\mu +1}$$ (cf. [F]).

\proclaim Theorem 2. The cohomology groups $H^2(\Diff ^+(S^1);{\cal
F}_\lambda )$, where $\lambda =0,1,2,5,7$ are generated by the following
non-trivial 2-cocycles:

\begin{eqnarray}
B_0(\Phi ,\Psi ) &= &const(\Phi ,\Psi ) =B(\Phi ,\Psi ) \\
B_1(\Phi ,\Psi ) &= &\Psi^{\star}(l\Phi )\cdot dl\Psi    \\
B_2(\Phi ,\Psi ) &= &\Psi^{\star}(l\Phi )\cdot S\Psi   \\
B_5(\Phi ,\Psi ) &= &\left| \begin{array}{cc} \Psi^{\star}S\Phi & S\Psi \\
\left(\Psi^{\star}S\Phi\right)^{\prime} & \left(S\Psi\right)^{\prime}
\end{array}\right| \\
B_7(\Phi ,\Psi ) &= &2\left| \begin{array}{cc} \Psi^{\star}S\Phi & S\Psi \\
\left(\Psi^{\star}S\Phi\right)^{\prime\prime\prime} &
\left(S\Psi\right)^{\prime\prime\prime}
\end{array}\right|-9\left| \begin{array}{cc}
\left(\Psi^{\star}S\Phi\right)^{\prime} & \left(S\Psi\right)^{\prime} \\
\left(\Psi^{\star}S\Phi\right)^{\prime\prime} &
\left(S\Psi\right)^{\prime\prime} \end{array}\right|\cr &&
-\frac{32}{3}(S\Psi +S(\Phi \circ \Psi ))B_5(\Phi ,\Psi )
\end{eqnarray}
\par

{\bf Remark}. The central extension of $\Diff ^+(S^1)$ by ${\cal F}_0\simeq
C^{\infty }(S^1)$ is in fact, a semi-direct product of the Bott-Virasoro
group by the module of functions: it is given by the Thurston-Bott cocycle.
\vskip 0,3cm

\subsection{Extensions of the Lie algebra $\Vect (S^1)$}

The following theorem classifies non-trivial extensions (2).

\proclaim Proposition 1. The cohomology group

$$H^2(\Vect (S^1);{\cal F}_\lambda )=\left\{\matrix{
{\bf R}^2 &, \lambda = 0,1,2  \cr
{\bf R} &, \lambda = 5,7  \cr
0&,\lambda \not = 0,1,2,5,7 \cr
}\right.$$

In other words, there exist 2 non-isomorphic non-trivial extensions of
$\Vect (S^1)$ by the module ${\cal F}_\lambda $ for $\lambda =1,2$; and an
unique non-trivial extension for each $\lambda =0,5,7$.
\vskip 0,3cm

Let us describe 2-cocycles on $\Vect (S^1)$ with values in ${\cal F}_\lambda
$ representing the non-trivial cohomology classes.

\proclaim Theorem 3. The cohomology groups $H^2(\Vect (S^1);{\cal F}_\lambda
)$, where $\lambda =0,1,2,5,7$ are generated by the following 8 non-trivial
2-cocycles:

\begin{eqnarray}
\bar c_0\left(f,g\right) &= &\left|
\begin{array}{cc} f&g \\
f^{\prime}&g^{\prime}
\end{array}\right| \\
c_0\left(f,g\right) &= &const(f,g)=\omega (f,g)\\
c_1\left(f,g\right) &= &\left| \begin{array}{cc}
f^{\prime}&g^{\prime} \\
f^{\prime\prime}&g^{\prime\prime}\\
\end{array}\right| dx\\
\bar c_1\left(f,g\right) &= & \left| \begin{array}{cc}
f&g \\
f^{\prime\prime}&g^{\prime\prime} \\
\end{array}\right| dx\\
c_2\left(f,g\right) &= &
 \left| \begin{array}{cc}
f^{\prime}&g^{\prime} \\
f^{\prime\prime\prime}&g^{\prime\prime\prime} \\
\end{array}\right| \left(dx\right)^2\\
\bar c_2\left(f,g\right) &=
&\left| \begin{array}{cc}
f&g \\
f^{\prime\prime\prime}&g^{\prime\prime\prime} \\
\end{array}\right| \left(dx\right)^2\\
c_5\left(f,g\right) &= & \left| \begin{array}{cc}
f^{\prime\prime\prime}&g^{\prime\prime\prime} \\
f^{\left(IV\right)}&g^{\left(IV\right)} \\
\end{array}\right| \left(dx\right)^5 \\
c_7\left(f,g\right)&= &
 \left(2\left| \begin{array}{cc}
f^{\prime\prime\prime}&g^{\prime\prime\prime} \\
f^{\left(VI\right)}&g^{\left(VI\right)} \\
\end{array}\right| -9 \left| \begin{array}{cc}
f^{\left(IV\right)}&g^{\left(IV\right)} \\
f^{\left(V\right)}&g^{\left(V\right)} \\
\end{array}\right| \right)\left(dx\right)^7
\end{eqnarray}

\par

Let us denote $g_i$ the Lie algebras given by the non-trivial cocycles
$c_i$ and $\bar g_i$ the Lie algebras given by the non-trivial cocycles
$\bar c_i$.
\vskip 0,3cm

{\bf Remark} 1. The algebra cocycles $c_0,c_1,c_2,c_5,c_7$ correspond to the
group cocycles $B_0,B_1,B_2,B_5,B_7$. The algebra cocycles $\bar c_0,\bar
c_1,\bar c_2$ can not be "integrated" to the group $\Diff ^+(S^1)$.
\hfill\break
2. The Lie algebra $g_0$ is a semi-direct product of the
Virasoro algebra by the module of functions
\hfill\break
3. The Lie algebra
 $g_5$ was considered in [OR].

\subsection{Central extensions}

Each of the constructed Lie groups and Lie algebras (except $G_0,g_0$), has
at least one non-trivial central extension given by prolongations of the
Thurston-Bott and Gelfand-Fuchs cocycles. Thus, one has 4 Lie groups
which are non-trivial extensions of the Virasoro group and 7 Lie algebras
which are non-trivial extensions of the Virasoro algebra, as described
through the following diagram.

$$\matrix{
&&&&0&&0&&\cr
&&&&\downarrow &&\downarrow &&\cr
&&&&{\cal F}_{\lambda }&=&{\cal F}_{\lambda
}&&\cr
&&&&\downarrow &&\downarrow &&\cr
0&\rightarrow &{\bf R}&\rightarrow &\widehat g_{\lambda
}&\rightarrow   &g_{\lambda }&\rightarrow &0\cr
&&\parallel &&\downarrow &&\downarrow &&\cr
0&\rightarrow &{\bf R}&\rightarrow &Vir&\rightarrow
&Vect(S^1)&\rightarrow &0\cr
&&&&\downarrow &&\downarrow &&\cr
&&&&0&&0&&\cr
}$$

An interesting fact is that the Lie algebras $\bar g_1$ and $g_1$ possess new
central extensions.
\vskip 0,3cm

Let us give here the complete list of non-trivial central extensions of
the  constructed Lie algebras.
\proclaim Proposition 2. Non-trivial central extensions of the Lie
algebras $g_i$ and $\bar g_i$ are given by the following list.\par

1)Each one of the Lie algebras $g_1,g_2,g_5,g_7$ and $\bar g_0,\bar g_1,\bar
g_2$ has a non-trivial
central extension given by :

$$c\left(\left(f,a\right),\left(g,b\right)\right)= \omega (f,g)$$
There exist two more non-trivial central extensions:
\vskip 0,3cm

2) A central extension of the Lie algebra $\bar g_1$ given by the cocycle:
\begin{eqnarray}
c\left(\left(f,a\right),\left(g,b\right)\right)=
\int_{S^1}(fb-ga)\,dx
\end{eqnarray}

3) A central extension of the Lie algebra $g_1$ given by the cocycle:
\begin{eqnarray}
c\left(\left(f,a\right),\left(g,b\right)\right)=
\int_{S^1}(f'b-g'a)\,dx
\end{eqnarray}

Let us consider also central extensions of a
semi-direct product: $\Vect (S^1)\triangleright {\cal F}_\lambda $.

\proclaim Proposition 3. The cohomology group
$$H^2(\Vect (S^1)\triangleright {\cal F}_\lambda ;\;{\bf R})=\left\{\matrix{
{\bf R}^3, \lambda =0,1  \cr
{\bf R},\lambda \not = 0,1 \cr
}\right.$$
\par

The cocycles which define central extensions of
$\Vect (S^1)\triangleright {\cal F}_0$ nonequivalent to the Virasoro
extension, are:

\begin{eqnarray}c\left(\left(f,a\right),\left(g,b\right)\right)=
\int_{S^1}(f''b-g''a)\,dx
\end{eqnarray}
and
\begin{eqnarray}c((f,a),(g,b))=\int_{S^1}(adb-bda)
\end{eqnarray}
One remarks, that the last cocycle defines the structure infinite-dimensional
Heisenberg algebra on the space ${\cal F}_0\simeq C^{\infty}(S^1)$. This
Lie algebra was considered in [ACKP].

In the case of $\Vect (S^1)\triangleright {\cal F}_1$, non-trivial cocycles are
also given by the formulae (17) and (18).

\section{Lie algebras $g_5$ and $g_7$ and Moyal bracket}
Consider the standard Poisson bracket on ${\bf R}^2$:
$$\{F,G\}=F_qG_p-F_pG_q$$
The Lie algebra of functions on ${\bf R}^2$ has a non-trivial
formal deformation which is called the Moyal bracket (see e.g. [FLS]).
\vskip 0,3cm

Consider the following bilinear operations invariant under the action of the
group $SL(2,{\bf R})$ of all linear symplectic transformations of ${\bf
R}^2$:

 \begin{eqnarray}
\{F,G\}_m=\sum_{i=0}^{m}(-1)^{i}
\left (\matrix{
m \cr
i \cr
}\right )
\frac {\partial ^mF}{\partial
^{m-i}q\partial ^{i}p}\frac {\partial ^mG}{\partial ^{i}q\partial
^{m-i}p} \end{eqnarray}
For example, $\{F,G\}_0=FG,\;\{F,G\}_1=\{F,G\}$,
$$\{F,G\}_2=F_{qq}G_{pp}-2F_{qp}G_{qp}+F_{pp}G_{qq}$$
$$\{F,G\}_3=F_{qqq}G_{ppp}-3F_{qqp}G_{qpp}+3F_{qpp}G_{qqp}-F_{ppp}G_{qqq}$$
etc.
\vskip 0,3cm

The Moyal bracket is defined as a formal series:
\begin{eqnarray}
\{F,G\}_t=\{F,G\}_1+{t\over 6}\{F,G\}_3+{t^2\over 5!}\{F,G\}_5+...
\end{eqnarray}

It verifies the Jacobi identity and defines a Lie algebra structure on the
space $C^{\infty }({\bf R}^2)[[t]]$ of formal series in $t$ with functional
coefficients.
\vskip 0,3cm

The relationship between the Lie algebras $g_5$, $g_7$ and the Moyal
bracket is based on the following realisation of tensor densities by
homogeneous functions on ${\bf R}^2$.

\subsection{Tensor densities on $S^1$ and homogeneous functions on ${\bf
R}^2$}

Consider homogeneous functions on ${\bf R}^2\setminus \{0\}$ (with
singularities in the origin): $F(\kappa q,\kappa p)=\kappa ^{\lambda
}F(q,p)$, where $\kappa >0$.

\proclaim Lemma 1. (i) The space of homogeneous functions of degree 2 on
${\bf R}^2\setminus \{0\}$ is a subalgebra of the Poisson Lie algebra
$C^{\infty }({\bf R}^2\setminus \{0\})$
\hfill\break
(ii) This subalgebra is isomorphic to $\Vect (S^1)$.\par
{\bf Proof}. The isomorphism is given by:
$f=f(x)d/dx\leftrightarrow F=r^2f(\phi )$
where $r,\phi$ are the polar coordinates.
\vskip 0,3cm
Moreover, the Poisson bracket defines a structure of $\Vect (S^1)$-{\it
module} on the space of homogeneous functions. Let $F$ and $G$ be
homogeneous functions of degree 2 and $\lambda $ correspondingly. Then, their
Poisson bracket $\{F,G\}$ is again a homogeneous function of degree $\lambda
$.

\proclaim Lemma 2. There exists an isomorphism of $\Vect
S^1$-modules: the space of homogeneous functions of degree $\lambda $ on
${\bf R}^2\setminus \{0\}$ and the space of tensor densities ${\cal
F}_{-\frac {\lambda }{2}}$.\par
{\bf Proof}. Take the mapping given by:

\begin{eqnarray}
f=f(x)(dx)^{-\frac {\lambda }{2}}\longleftrightarrow F=r^{\lambda }f(\phi )
\end{eqnarray}
It is
easy to verify that the Poisson bracket corresponds to the Lie derivative
by this mapping: $L_fg\longleftrightarrow \{F,G\}$.

\proclaim Corollary. (i) There exists a series of $SL_2$-invariant
operations  $$\{.,.\}_n:{\cal F}_{\lambda }\otimes {\cal F}_{\mu
}\rightarrow  {\cal F}_{\lambda + \mu +n}$$
on the space of tensor densities on $S^1$.
\hfill\break
(ii) Moreover, there exists a Lie structure given by the Moyal bracket (21)
on the space of tensor densities on $S^1$.\par

{\bf Remark}. The isomorphism (23) is in fact, much more general. It is
valid in the case on an arbitrary contact manifold (see [OR]).

\subsection{Cocycle $c_5$}

Let us substitute two vector fields $f=f(x)/dx,\;g=g(x)d/dx$
(corresponding to homogeneous functions of degree 2 on
${\bf R}^2\setminus \{0\}$) to
the Moyal bracket. We get a formal series in $t$ with coefficients in the
space of tensor densities.

\proclaim Lemma 3. (i) If $f,g\in \Vect (S^1)$, then $\{f,g\}_3=\{f,g\}_5\equiv
0$.
\hfill\break
(ii) The first non-zero term: $\{f,g\}_7\in {\cal F}_5$ is proportional to
$c_5$.\par

{\bf Proof}. Let $F,G$ be two functions on ${\bf R}^2\setminus \{0\}$,
then $\{F,G\}_7$ is a homogeneous function of degree $-10$. Thus,
$\{f,g\}_7\in {\cal
F}_5$. It is easy to verify that $\{f,g\}_7=20160c_5$.
\vskip 0,3cm

Consequently, $c_5$ is a 2-cocycle. Indeed, the Jacobi identity for the
Moyal bracket implies that the first non-zero term in this series is a
2-cocycle on  $\Vect (S^1)$.

\subsection{Cocycle $c_5$}
\proclaim Lemma 4. The second non-zero term of the Moyal bracket:
$\{f,g\}_9\in {\cal F}_7$ is proportional to
$c_7$.\par
{\bf Proof}: straightforward.
\vskip 0,3cm
Let us prove that $\{f,g\}_9$ is again a 2-cocycle on $\Vect (S^1)$.
The Jacobi identity for the Moyal bracket $\{f,g\}_t$ implies:

$$\{f,\{g,h\}_9\}_1+\{f,\{g,h\}_1\}_9+\{f,\{g,h\}_7\}_3\;\;(+cycle)=0$$
for any $f,g,h\in \Vect (S^1)$. One checks that the expression
$\{f,\{g,h\}_7\}_3$ is proportional to $f'''(g'''h'^{\vee }-g'^{\vee
}h''')$, and so one gets: $\{f,\{g,h\}_7\}_3\;\;(+cycle)=0$. We obtain the
following relation:

$$\{f,\{g,h\}_9\}_1+\{f,\{g,h\}_1\}_9\;\;(+cycle)=0$$
which means that $\{f,g\}_9$ is a 2-cocycle. Indeed, recall that for any
tensor density $a$, $\{f,a\}_1=L_fa$. Therefore, this relation coinsides
with:

$$L_f\{g,h\}_9+\{f,\{g,h\}_1\}_9\;\;(+cycle)=0$$
which is exactly the relation $d\{.,.\}_9=0$.

\subsection{Group cocycle $B_7$}
Let us recall that the mapping $s:f(x)d/dx\rightarrow f'''(x)(dx)^2$ is a
2-cocycle (an algebraic analogue of the Schwarzian derivative) generating the
cohomology group $H^1(\Vect (S^1);{\cal F}_2)$.  It is easy to check, that
the following relation is satisfied:
$$\{f,g\}_9=504\{f''',g'''\}_3$$

Thus, it is natural to look for a group version of the cocycle $c_7$ in the
following form: $B(\Phi,\Psi)=\{\Psi ^*S(\Phi ),S(\Psi )\}_3$. However,
this formula does not define a cocycle on the group $\Diff ^+(S^1)$ since the
operation $A\otimes B\mapsto \{A,B\}_3$ is not invariant.

\proclaim Lemma 5. Let $a,b\in {\cal F}_2$, $\Phi $ is a diffeomorphism, then
$$\{\Phi ^*a,\Phi ^*b\}_3=\Phi ^*\{a,b\}_3+480S(\Phi )\Phi ^*\{a,b\}_1$$
where $S(\Phi )$ is the Schwarzian derivative\par

{\bf Proof}: straightforward.
\vskip 0,3cm

Let us verify now that $B_7$ is a 2-cocycle. One must show that
$$dB_7(\Phi ,\Psi ,\Xi )=B_7(\Phi ,\Psi \circ \Xi )-B_7(\Phi \circ \Psi
,\Xi )+B_7(\Psi ,\Xi )-\Xi ^*B_7(\Phi ,\Psi )=0$$
for any $\Phi ,\Psi ,\Xi \in \Diff ^+(S^1)$.
\vskip 0,3cm

Let us take the expression: $\bar B(\Phi ,\Psi )=\{\Psi ^*S(\Phi ),S(\Psi
)\}_3$. Lemma 5 implies: $d\bar B(\Phi ,\Psi ,\Xi )=480S(\Xi )\cdot \Xi
^*\{\Psi ^*S(\Phi ),S(\Psi )\}_1$. Consider anoter expression $\widetilde
B(\Phi ,\Psi )=(S(\Psi )+S(\Phi \circ \Psi ))\{S(\Phi ),S(\Psi )\}_1$. A
simple computation gives: $d\widetilde B(\Phi ,\Psi ,\Xi )= 3S(\Xi )\cdot
\Xi ^*\{\Psi ^*S(\Phi ),S(\Psi )\}_1$. Thus, the expression
$$B(\Phi ,\Psi )=\{S\Phi ,S\Psi \}_3-160(S\Psi +S\Phi \circ \Psi )\{S\Phi
,S\Psi \}_1$$
is a 2-cocycle. One checks that this formula is proportional to the formula
(8).

\section{Proofs of main theorems}

The proofs are quite simple but they use high technics of cohomology of
infinite-dimensional Lie algebras. All the necessary details can be found in
the book of D.B.Fuchs.
\vskip 0,3cm

{\bf Proof of Proposition 1}

Cohomology groups of Lie algebras of smooth vector fields with coefficients
in modules of tensor fields, where calculated essentially in [GF 2], [T]
(see also [F], p.147). Let us recall here the answer.
\vskip 0,3cm

The cohomology group

$$H^q(\Vect (S^1);{\cal F}_{\frac {3r^2\pm r}{2}})=H^{q-r}(Y(S^1);{\bf R} )$$
where $Y(S)\simeq S^1\times S^1\times \Omega S^3$,
and $H^q(\Vect (S^1);{\cal F}_{\lambda })=0$ if $\lambda \not =0,1,2,5,7,...,
\frac {3r^2\pm r}{2},...$.
The cohomology ring $H^*(Y(S^1);{\bf R} )$ is a free anti-commutative
algebra with generators in dimensions 1,1,2.
\vskip 0,3cm

One obtains immediatelly the proof of Proposition 1.
\vskip 0,3cm

{\bf Proof of Theorem 3}

The cohomology ring $H^2(\Vect (S^1);C^{\infty }(S^1))$ is generated by three
cocycles:
$$f(x){d\over dx}\mapsto f(x),\;\;\;\;f(x){d\over dx}\mapsto f'(x),\;\;\;\;
\omega (f,g)$$
(cf. Theorem 2.4.12 of [F]).
Thus, one has two non trivial cocycles (9),(10).

\vskip 0,3cm

$H^2(\Vect (S^1);{\cal F}_1)$ is a free $H^2(\Vect (S^1);C^{\infty
}(S^1))$-module. This fact implies formulae (11), (12) for the generating
cocycles. \vskip 0,3cm

The cocycles (13) and (14) can be obtained from the isomorphism: ${\cal
F}_2\simeq {\cal F}_1\otimes {\cal F}_1$.
\vskip 0,3cm

The proof that formulae (15) and (16) define 2-cocycles on $\Vect (S^1)$ is
given in the sect.3. (For the cocycle (15) it follows also from the formula
(7) for the group cocycle). These cocyles define non-trivial cohomology
classes. Indeed, one can check, that the Lie algebras $g_5$ and $g_7$ do not
verify the same identities as $\Vect (S^1)$.
\vskip 0,3cm

{\bf Proof of Theorem 1}

The Van Est cohomology ring for the Lie group $\Diff ^+(S^1)$ is defined by
the following isomorphism (cf [F], p.244):
$$H^*_c(\Diff ^+(S^1); \;{\cal F}_{\lambda })\simeq H^*(\Vect (S^1), S0(2);\;
{\cal F}_{\lambda })$$
(where $SO(2)\subset \Diff ^+(S^1)$ is the maximal compact subgroup of
"rotations" of $S^1$). Thus, $H^*_c(\Diff ^+(S^1); \;{\cal F}_{\lambda })=0$ if
$\lambda \not =0,1,2,5,7$.

\vskip 0,3cm

The cohomology ring $H^*(\Vect (S^1), SO(2);\;
{\cal F}_{\lambda })$ is defined by cochains which are identically zero on
the subalgebra $so(2)\subset \Vect (S^1)$ (that means, by cochains given by
differential operators without zero order terms). To
prove that the cocycles $\bar c_0, \bar c_1, \bar c_2$ can not be integrated
to $\Diff ^+(S^1)$, one must show now that the cohomology classes of these
cocycles can not be represented by such cocycles.
\vskip 0,3cm

Suppose, that there exists a cocycle $\bar c_0'$ cohomological to $\bar
c_0$ such that $\bar c_0'(f,g)=\sum_{i,j\geq 1}c_{ij}f^{(i)}g^{(j)}$. Then
$\bar c_0-\bar c_0'$ is a differential of some 1-cochain $\sigma $. But it
is easy to verify that the expression $d\sigma (f,g)$ depends only on the
derivatives of $f$ and $g$. The contradiction means that there is no
cohomology class in $H^*(\Vect (S^1), SO(2);\;
{\cal F}_0)$ corresponding to the cocycle $\bar c_0$. Thus, $H^2(\Vect (S^1),
SO(2);\; {\cal F}_0)=0$. Analogous arguments are valid for  $\bar
c_1, \bar c_2$.
\vskip 0,3cm

To finish the proof of the theorem, one should show that the cocycles
$c_1,c_2,c_5,c_7$ correspond to some group cocycles, but it follows from
formulae given by Theorem 2.

\vskip 0,3cm

{\bf Proof of Theorem 2}

It follows from the construction, that the mappings (4)-(7) are cocycles on
$\Diff S^1$. It was proved in the sect.3 that the formula (8) defines a
cocycle. The Lie algebra cocycles associated with the group cocycles
(4)-(8), are  $c_0,c_1,c_2,c_5,c_7$ correspondingly. This proves that the
cocycles (4)-(8) represent non-trivial cohomology classes.

\vskip 0,3cm

{\bf Proof of Propositions 2 and 3}

In general, let $L$ be a Lie algebra and $M$ is an $L$-module, consider a
Lie structure $L_M$ on $L\oplus M$ with the commutator:
$$[(X,\xi )(Y,\eta )]=([X,Y],X(\eta )-Y(\xi )+\alpha (X,Y))$$
where $\alpha $ is a 2-cocycle: $\alpha \in Z^2(L;M)$. Let $c$ be a
2-cocycle on $L_M$ with scalar values. Then:

1) The restriction $c\mid _{L\otimes L}$ is a 2-cocycle on $L$.

2) The condition $dc(X,Y,Z)=0$ implies: $c(\alpha
(X,Y),Z)\;(+cycle)=0$.

3) $dc(X,a,b)=0$ implies:
$c(X(a),b)+c(a,X(b))=0$. That is, the restriction $\widetilde c=c\mid
_{M\otimes M}$ is L-invariant.

4) $dc(X,Y,a)=0$ means:
$$c(\alpha (f,g),a)=c(X(a),Y)-c(Y(a),X)-c([X,Y],a)$$
Thus, the linear mapping $\bar c:L\rightarrow M^*$ given by the relation:
$\bar c(X)a=c(X,a)$
satisfies the following condition:
$$d\bar c =\widetilde c\circ \alpha $$

Let us apply these facts to the case: $L=\Vect (S^1),\; M={\cal
F}_{\lambda }$.
\vskip 0,3cm

First of all, the restriction $c\mid _{L\otimes L}$ is proportional to the
Gelfand-Fuchs cocycle.
\vskip 0,3cm

The unique invariant bilinear mapping $\widetilde c:{\cal F}_{\lambda
}\wedge {\cal F}_{\lambda }\rightarrow {\bf R}$ is
$$\widetilde c(a,b)=\int_{S^1}(adb-bda)$$
if $\lambda=0$. One obtains the cocycle (20) for the semi-direct product.
This cocycle can not be extended on the Lie algebra $g_0$, since the
property  $d\bar c =\widetilde c\circ \alpha $ is not satisfyied. If
$\lambda \not =0$, then  $\bar c$ is 1-cocycle.

\vskip 0,3cm
The dual module ${\cal F}_{\lambda }^*$ is isomorphic to ${\cal
F}_{1-\lambda }$.
Consider the cohomology group: $H^1(\Vect (S^1); {\cal F}_{1-\lambda })$,
where $\lambda = 0,1,2,5,7$. It is not trivial in two cases: $ \lambda =
0,1$. Otherwise, there is no non-trivial extensions which are not
equivalent to the Virasoro one. The group $H^1(\Vect (S^1); {\cal F}_0)$ is
generated by the following two elements:

$$f(x)d/dx\mapsto f(x)\;\;\;\;f(x)d/dx\mapsto f'(x)$$
 (the first part of Theorem 2.4.12 of [F]). One
obtains the cocycles (17) and (18) which satisfy the condition 2). The group
$H^1(\Vect (S^1)\Vect (S^1); {\cal F}_1)$ has one generator
$$f(x)d/dx\mapsto f''(x)dx$$
(not $f\mapsto f'dx$: there is a misprint in [F] here). One gets the cocycle
(19).

\vfill\eject

{\bf References}
\vskip 0.5cm
[ACKP] Arbarello E., De Concini C., Kac V.G., Procesi C. Moduli Space
of Curves and Representation Theory, Comm. Math. Phys., 117 (1988) 1-36.
\vskip 0.3cm
[B] Bott R. On the characteristic classes of groups of diffeomorphisms,
Enseign Math. 23:3-4 (1977), 209-220.
\vskip 0.3cm
[Br] Brown K.S. Cohomology of Groups, Springer-Verlag, New York,
Heidelberg, Berlin 1982.
 \vskip 0.3cm
[FLS] Flato M., Lichnerowicz A., Sternheimer D. D\'eformations
1-differentiables des alg\`ebres de Lie attach\'ees \`a une vari\'et\'e
symplectique ou de contact, Compositio Math. 31 (1975) 47-82.
\vskip 0,3cm
[F] Fuchs D. B., Cohomology of infinite-dimensional Lie
algebras, Consultants Bureau, New York, 1987.
\vskip 0.3cm
[GF 1] Gelfand I. M., Fuchs D. B. Cohomology of the Lie algebra of vector
fields on the circle, Func. Anal. Appl. 2:4 (1968), 342-343.
\vskip 0.3cm
[GF 2] Gelfand I. M., Fuchs D. B. Cohomology of the Lie algebra of vector
fields with non-trivial coefficients, Func. Anal. Appl. 2:4 (1970) 10-25.
\vskip 0.3cm
[OR] Ovsienko V. Yu., Roger C., Deformations of Poisson brackets and
extensions of Lie algebras of contact vector fields, Russian Math. Surveys
47:6 (1992) 135-191.
\vskip 0.3cm
[T] Tsujishita T. On the continuous cohomology of the Lie algebra of
vector fields, Proc. Jap. Acad. A53 ,N.4 (1977), 134-138.

\end{document}